# Software Process Measurement and Related Challenges in Agile Software Development: A Multiple Case Study

Prabhat Ram[1], Pilar Rodriguez[1], and Markku Oivo[1]

[1] M3S, University of Oulu, 90014 Oulu, Finland
(prabhat.ram, pilar.rodriguez, markku.oivo)@oulu.fi

**Abstract.** Existing scientific literature highlights the importance of metrics in Agile Software Development (ASD). Still, empirical investigation into metrics in ASD is scarce, particularly in identifying the rationale and the operational challenges associated with metrics. Under the Q-Rapids project (Horizon 2020), we conducted a multiple case study at four Agile companies. We used the Goal Question Metric (GQM) approach to investigate the rationale explaining the choice of process metrics in ASD, and challenges faced in operationalizing them. Results reflect that companies are interested in assessing process aspects like velocity, testing performance, and estimation accuracy, and they prefer custom metrics for these assessments. Companies use metrics as a means to access and even capitalize on the data, erstwhile inaccessible due to technical or process constraints. However, development context of a company can hinder metrics operationalization, manifesting primarily as unavailability of the data required to measure metrics. The other challenge is the uncertain potential of metrics to help derive actionable inputs to facilitate decision-making. Essentially, development context has a strong influence over a company's choice of process metrics, rationale, and challenges to operationalize these metrics.

## 1 Introduction

Software measurement enables understanding of cost and quality of software development [1], and it supports in planning, monitoring, controlling, and evaluating software processes [2]. The increasing popularity of Agile Software Development (ASD) [3, 4] makes understanding of software metrics in agile context more relevant. Research recognizes the need for agile organizations to use metrics, but empirical research on metrics in industrial ASD remains scarce [3, 5].

Existing studies discuss use of metrics in ASD for planning and tracking software development [5], estimating effort [6], understanding development performance and product quality [3], and reporting ASD progress and quality to stakeholders not involved in the actual development [7]. These studies propose metrics focused on specific quality improvement goals, but most of them present initial emerging results that have not been evaluated within larger industrial context. Kupiainen et al. [5] conducted a systematic literature review to investigate the reasons and actual use of metrics in ASD, and drew a similar conclusion that there is a lack of empirical studies in industrial context. There is also a need to investigate metrics that can influence process improvement in ASD, how they are operationalized [5], and the accompanying challenges. These gaps motivate the following research questions (RQ):

- **RQ1:** What metrics are software-intensive companies interested in to assess their ASD processes, and the rationale behind them?
- **RQ2:** What are the challenges faced by software-intensive companies in operationalizing the metrics to assess their ASD processes?

We conducted our research in the context of the Q-Rapids[1] project, a Horizon 2020 (H2020) project, involving three research organizations and four ICT companies. The goal of the project is to develop an agile-based, data-driven, quality-aware rapid software development framework [8]. Building on top of the project progress [9], we used Goal Question Metric (GQM) to collect data at the four case companies, and facilitate the multiple case study to help answer the research questions. We conducted 12 GQM workshops (three with each case company), involving a total of 19 practitioners. These case companies have several years of experience in ASD, are of different size, and focus on diverse industrial domains. Differences in development contexts at the four case companies enabled rich data collection and comparative analysis of our findings.

In comparison to existing literature, our study contributes in the following ways:

- We present empirical evidence on the metrics that software-intensive companies use to assess their ASD process.
- We identify and discuss aspects that influence their choice of metrics.
- We draw a metric-centric comparison among the four case companies, and discuss the challenges faced in operationalizing these metrics.

The remainder of the paper is structured as follows: Section 2 covers background and related work on the topic. Section 3 describes the research methodology, followed by the multiple case study findings in Section 4, and discussion in Section 5. Section 6 presents the threats to validity to our paper, followed by conclusion and future research directions in Section 7.

## 2      Background and Related Work

Software measurement in ASD is different from traditional software development methods, primarily because of the differences in processes in these methods [5, 10]. Some of the metrics defined for ASD include velocity, software size estimation, burn-down chart, cumulative flow, etc. [10]. Systematic reviews have investigated the state of the art on the use and impact of ASD metrics in industry [5], to provide an overview of metrics on effort estimation [11], as well as effort estimation practices in agile, iterative, and incremental software development [12]. Although the scientific literature has discussed the role metrics play in ASD, there is still insufficient empirical evidence on that role in industrial context, especially in view of large companies [13].

---

[1] http://www.q-rapids.eu/

The systematic review by Kupiainen et al. [5] investigated the state of the art on the industry use of metrics in ASD, rationale, and the consequent impact. The authors argue that the use of metrics in ASD and traditional software development companies is similar, as the emphasis in both appear to be on planning and monitoring. In addition to the metrics described in Agile literature, companies also use custom metrics to measure aspects such as business value, defect count, and customer satisfaction. Kupiainen et al. [5] call for more empirical studies to investigate rationales behind metrics use in ASD.

Most primary studies (case studies) in Kupiainen et al. [5] investigate the impact of using Agile in a software company, and use metrics as a tool to measure that impact. Very few studies have enquired the role of metrics in ASD, as used by practitioners. For example, Dubinsky et al. [14] reported on the experience of using software metrics program at an XP development team of the Israeli Air Force. The authors found that using metrics to measure the amount, quality, pace, and work status could lead to accurate and professional decision-making. However, authors focused mainly on the impact of metrics use, and not on the challenges in operationalizing them. A similar approach was followed by Díaz-Ley et al. [15], where a measurement framework, customized for SMEs, was applied in an industrial context. One key benefit the authors reported was being able to define better measurement goals that align with the company's maturity. However, similar to [14], the authors did not discuss the challenges of operationalizing metrics. Furthermore, the focus on process metrics was also missing from these studies.

The research gaps identified in the systematic literature review conducted by Kupiainen et al. [5] serve as the foundation for the research questions we address in our paper. Similar to [15], we adopted GQM approach to gather empirical evidence about process metrics from diverse companies using Agile development practices, and identify rationales and challenges in operationalizing these metrics. The Goal-Question-Metric (GQM) approach establishes a mechanism to define and interpret software measurement driven by organizational goals. GQM helps specify a goal to be measured and refined into a set of quantifiable questions. These questions, in turn, help define a set of metrics and data for collection [16]. All of this information is recorded on an "abstraction" sheet to provide a structured approach [17] in data interpretation. An abstraction sheet is a tool used to record interviews in the GQM approach. Essentially, GQM approach helps trace a goal to the data that can define that goal operationally, and provide a framework to interpret that data with respect to the goal [3]. Linking data (metrics answering the questions) to goals ensures that relevant metrics are collected, allowing for control over what is collected and its quality [16].

## 3 Research Methodology

Following the case-study guidelines by Runeson et al. [19], this paper reports a multiple case study involving four case companies. We used GQM for data collection, followed by thematic synthesis [20] for analyzing the results, and construct themes to answer the research questions.

## 3.1 Research Context

Table 1. presents the development context for the four case companies. The development context is distinguished across company size, software development method, and development team composition involved in our case study.

Table 1. Case Study Context

| Company & Size | Product | Development Method | Development Team |
|---|---|---|---|
| A – Medium | Production Testing Software Framework | Scrum | One team with 6-7 members |
| B – Large | Software platform | Scrum | Eight globally distributed sub-teams |
| C – Large | Software modeling tool | Ad-hoc process following Agile principles. | One team with 9 members |
| D - Small | "R" Project | ScrumBan & Scrumbut | 10 members with same core team |

*Company A* is a medium-size company (over 600 employees) that develops secure communication and connectivity solutions for multiple industry domains. In a bid to achieve efficiency and shorten time-to-market, the company moved to agile and lean software development about 10 years ago. The company aims to develop metrics to measure ASD process to introduce high-level transparency, and a more data-driven and evidence-based decision-making process. The software development process is mainly Scrum-based. For our study, we worked with one of the software teams developing a hardware-testing framework, used internally to test secure solutions that Company A develops.

*Company B* is a large-size company (over 100,000 employees) developing distributed systems in telecommunication networks. The company aims to have a standardized way of working and tools to identify, analyze, manage, and implement quality requirements right until individual product releases. The software development method, at team level, is Scrum-based. The company's development unit is divided into multiple teams, which are further divided into sub-teams. We focused on the metrics the eight sub-teams will use to develop a software platform, which will be used to build other products.

*Company C* is a large-size company (over 900 employees) that develops a modeling tool used by developers for model driven development. The product is mature, with multiple releases already in the market. Company C wants to improve quality of the ASD process through early detection of anomalies in development. The company does not follow any formalized method like Scrum, but uses different software development methods that adhere to agile principles. Thus, the company has defined their own agile way of working. They engage in iterative development, but do not have any pre-defined sprint cycles, as they focus more on current issues.

*Company D*, a small company, is engaged in developing independent software products for multiple industrial domains. The company is targeting for metrics that allows its developers to anticipate design issues, security issues, and platform limitations. The core development team remains the same, but other team members may change from project to project. The software development process is Scrum-based, but with some exceptions. Here, the company uses a pre-software development process to acquire functional and quality requirements. Initial mock-ups and user stories collected during this process serves as the basis for the implementation process. The company develops software in iterations, and uses Kanban board to monitor the status of backlog items.

### 3.2 Data Collection

We collected data by conducting 12 GQM workshops, three with each case company. The GQM goal driving the workshops was, "*to analyze the agile/rapid software development process for the purpose of monitoring with respect to process performance/quality from the viewpoint of the process users in the context of the company case*". Based on this generic goal, questions were devised during the GQM workshops, and metrics to provide answers to those questions were elicited. We requested that individuals (stakeholders) involved in process management activities attend these workshops. Taking together all the four case companies, a total of 19 practitioners participated in the workshops. Participants included project managers, product owners, quality managers, and developers. Table 2 shows the details of these workshops:

Table 2. GQM Workshops

| Sessions | Parameters | Company A | Company B | Company C | Company D |
|---|---|---|---|---|---|
| Session 1 | Role (# participants) | Quality Lead, Developers (2), Requirement & Process Lead | Quality Manager (2), Project Manager, Developer (3) Development Manager (2 | Architect/Developer, Project Manager, R&D Manager, CEO / Product Owner | Product Owner, Project Manager, System Designer |
| | Data Collection | Documented | Both | Both | Documented |
| | Length (hrs) | 3.5 | 3.5 | 3 | 3 |
| Session 2 | Role (# participants) | Quality Lead, Developers (2), Requirement & Process Lead | Quality Manager, Project Manager, Developer (2), Development Manager | Architect/Developer and R&D Manager | Product Owner, Project Manager |
| | Data Collection | Both | Both | Both | Documented |
| | Length (hrs) | 2.5 | 2 | 1 | 1.5 |
| Session 3 | Role (# participants) | Quality Lead, Developers (2), Requirement & Process Lead | Quality Manager, Project Manager, Developer (2) | Architect/Developer | Project Manager |
| | Data Collection | Both | Both | Documented | Documented |
| | Length (hrs) | 2.5 | 1 | 0.5 | 1.5 |

Instead of starting from scratch, we built on top of a preliminary set of process metrics that were already identified by the case companies during earlier project tasks [9]. One of the first tasks in the project was to develop a 'Quality Model' that helps the case companies better define their understanding of quality [8, 9]. Software quality workshops were held to define this Quality Model, consisting of application of GQM+Strategies™, Quamoco, and GQM. Metrics were classified as either 'product factors' or 'process factors' based on whether they signified product or process characteristics [9]. Some process metrics emerged as relevant during these workshops, conducted between December 2016 and February 2017. However, only the metrics that could help the case companies assess and improve product quality were developed further, as that was the focus of the project's tasks. The process metrics that remained became the preliminary set to start the subsequent GQM workshops, conducted as part of this multiple case study between November 2017 and January 2018.

From the preliminary set, case companies chose the process metrics they considered relevant to their development context. Similarly, they discarded metrics from the preliminary set that they considered irrelevant for measuring Agile process performance and quality, and added new metrics that particularly focus on assessing their ASD processes. While eliciting metrics, we enquired participants to focus on details such as why the metric was relevant, how the metric would be measured (e.g. formula to measure the metric), data sources needed to obtain the data, and ways to operationalize the metric.

Of the 12 workshop sessions, we both recorded and documented seven sessions, and the rest could only be documented. Three researchers participated in the kick-off workshop with each case company. Two researchers participated in the subsequent workshops, where one conducted the workshop and the other documented it.

### 3.3    Data Analysis

Data collected during the GQM workshops consisted of the metrics recorded in GQM abstraction sheets [17], recordings, and supporting notes. Data was analyzed incrementally. At the end of every GQM session, one researcher analyzed the documentation/recording, and shared the analysis with the other researcher for corroboration. Next, before the following session in a case company took place, the analysis was shared with the individual case company for validation and feedback. This analysis helped answer the first part of RQ1 (choice of process metrics).

We used thematic synthesis to analyze the data (recordings, meeting minutes, and feedback) generated from the GQM workshops to answer the second part of RQ1 (metrics rationale) and RQ2 (operationalization challenges). One researcher performed line-by-line coding of the data and recorded concepts focusing on metrics rationale, and on concepts related to operationalization challenges. The researcher further analyzed the inductively coded concepts at a higher abstraction level to develop descriptive themes. Themes help transform large number of codes into a smaller analytical unit. Next, the researcher mapped these descriptive themes based on their interrelationships to develop higher-order analytical theme(s) for metrics rationale and for operationalization challenges. The thematic synthesis framework and the resulting themes were discussed with other researchers and refined further.

## 4 Results

Overall, we found that case companies targeted very similar process aspects, but adopted custom metrics to assess those process aspects. They also share very similar rationale and challenges in operationalizing these metrics.

### 4.1 RQ1: What metrics are software-intensive companies interested in to assess their ASD processes, and the rationale behind them?

A total of 132 metrics were elicited from the workshops (including metrics from the software quality workshops), available in Appendix A (https://goo.gl/nf1WLJ). For brevity, we present our results in Table 3 from factors' point of view, further elaborated in Appendix B (https://goo.gl/8zbScQ). 'Factors' here is a generalization of 'product factors' and 'process factors', and our focus is on the latter.

**Table 3.** Metrics and Rationale

| Factors | Measures… | Rationale |
|---|---|---|
| Testing Performance ** | …testing phase performance aspects like execution time | Track improvements & bottlenecks |
| Issues' Velocity ** | … capability to fulfil issues planned for a sprint | Assess & improve planning capability, Identify bottlenecks, Knowledge sharing |
| Code Quality * | … impact of code changes in source code quality | Knowledge sharing |
| Issues' Estimation Accuracy ** | …difference between effort estimated and actual effort invested for an issue | Resource planning |
| Testing Status * | ….unit test success density | Process improvement |
| Blocking Code * | …number of files not violating quality rule, which otherwise may block the flow of other coding activities | Identify bottlenecks |
| Delivery Performance *** | ….capability for on-time delivery, considering resource management | Identify bottlenecks |
| External Quality *** | …quality of a product from customer standpoint | Process improvement |
| Development Speed *** | …daily build progress | Data availability |
| Quality Issues' Specification * | …amount of issues entering backlog in an incomplete state | Traceability |

\* - Factors identified and completely defined in the software quality workshops [9]
\*\* - Factors identified during the software quality workshops [9], but metrics elicited/refined in the GQM workshops
\*\*\* - New factors identified and defined in the GQM workshops

A total of 10 factors were identified, of which seven make clear references to process aspects (*Testing Performance, Issues' Velocity, Issues' Estimation Accuracy, Testing Status, Delivery Performance, Development Speed,* and *Quality issues' specification*). *Code Quality* [9] and *External Quality* can be argued as more product-oriented factors. However, case companies argued that improvement in these factors can indicate good process, and so they should be considered from process-improvement standpoint as well.

We further describe the results for RQ1 based on the degree of commonality in choice of factors (and by extension, metrics) among the case companies.

**Factors common to all companies.** At factors level, all the case companies were interested in assessing *Testing Performance, Issues'*[2] *Velocity, Code Quality,* and *Issues' Estimation Accuracy*. For assessing *Testing Performance,* companies preferred mostly custom metrics. Company A defined metrics like 'Error leakage' and 'Average number of iterations in the code-review phase', aiming to track improvements in testing process and bottlenecks in review phases, respectively. Targeting code-review phase, Company B defined metrics like 'Actual feedback time from CI to developers' to assess *Testing Performance*. Similarly, Company C defined metrics like '% errors identified during a validation for a given release', but with the objective of learning what went wrong in testing and why, evident from the following quote: "At the end of the project…we may have to have an analysis about how the [development] process goes…what went wrong, or good, and why". Company D was interested in metrics that could meet three distinct objectives of informing developers about their progress, assisting Product Owner in development team management, and informing the management about the overall project status. The objective of keeping developers in the loop, in addition to tracking bottlenecks in testing, is reflected in their choice of custom metrics like 'No. of tickets that are pending tests' and 'No. of tickets in the "Ready" column' to assess *Testing Performance*.

For assessing *Issues' Velocity,* case companies A and C use several common process metrics like 'Average speed to resolve issues' and 'No. of issues/tickets/story points at start of the sprint'. These metrics will help Company A access the data their system is already producing, and even assess sprint planning capability. The latter rationale is reflected in one of the stakeholders' quote, "Do we allocate too much story points to a sprint?" For Company C, these metrics can help identify bottlenecks in releasing on time, reflected in the following quote, "…[focus was] not the performance of the process but the efficiency of the process. And the idea is to have products or projects on time with acceptable quality". Both Company B and C relied mainly on custom metrics to assess *Issues' Velocity*. Company B defined 11 custom metrics like 'No. of issues/tickets at start of the sprint' and 'No. of done issues at the end of the sprint' to track its sprint progress. Similarly, Company D defined nine custom metrics like 'No. of Ready issues', and 'No. of issues that are delayed' to learn if their sprint planning needs improvement.

As described in [9], in assessing *Code Quality,* three case companies used largely the same set of metrics. The metric of 'Complexity' was the only common metric for all the case companies. In contrast, Company D preferred mainly custom metrics like 'Code reliability', 'Code maintainability', 'Code security', etc. to assess *Code Quality*. Although not process oriented, these metrics can be used to measure process performance, as good quality process results in good quality code.

Lastly, all the case companies were interested in assessing *Issues' Estimation Accuracy*. The common rationale was to measure the accuracy with which a case company plans the effort (man-days) required to implement an issue. The fundamental metrics like 'Estimated effort of an issue/story point' and 'Real invested effort of an issue/story point' were common to all the case companies. However, they expressed difficulties when enquired about their plans to operationalize these metrics, which we elaborate upon in Section 4.2.

---

[2] A JIRA terminology that could represent a software bug, a project task, a helpdesk ticket, etc. - https://goo.gl/vNQGJE

**Factors common to three or two case companies.** *Testing Status* was common to three case companies (Company C being the exception). Only Company A and Company B shared similarities at metrics level, while Company D relied exclusively on custom metrics to assess this process factor.

*Blocking code* was common to Company B and C, and was assessed using the only metric '% files without critical/blocker quality rules'. This process factor refers to the condition where a particular file violates a predefined quality rule, which could block other coding activities [9]. A clear rationale could not be gathered for this metric, but it appears to satisfy Company C's aim to identify causes for delays in product releases.

**Custom Factors.** The factors of *Delivery Performance* and *External Quality* were exclusive to Company C, thereby requiring custom metrics to assess them. Based on the principle that process dictates product quality, *External Quality* comprises metrics that highlight product-related concerns resulting from process inefficiencies. Company C is interested in recording product quality related issues raised by the end users, map these to corresponding development processes, and improve upon them for subsequent product releases. Metrics under the *Delivery Performance* factor align well with the case company's aim to identify reasons for delay in releases.

Company A's interest in process metrics for assessing *Development Speed* was driven by the need to retrieve data from its Continuous Integration/Development (CI/CD) engine to measure their daily build performance. This decision finds support in their rationale of accessing and using the rich data, erstwhile inaccessible due to lack of appropriate retrieval mechanisms, which their system is producing.

Only Company B expressed interest in assessing *Quality Issues' Specification*, using the only metric '% issues completely specified'. The metric refers to the issues that have been completely specified in the backlog, and can, hence, commence implementation. Company B works on large-size features, with implementation spanning multiple teams and sub-teams using diverse tools. A given feature may be specified differently in different tools, making it a time-consuming task to trace these different specifications back to the original feature. It is this traceability that the chosen metric is expected to help with, by linking feature information across different tools with different specifications.

Overall, the four case companies were interested in measuring same factors (*Testing Performance, Issues' Velocity, Code Quality,* and *Issues' Estimation Accuracy*). Despite the commonality at factor level, these companies preferred mainly custom metrics for assessing them. This decision was dictated by the development context, especially the technical infrastructure available and the existing software development process.

The distinct rationales of planning, tracking improvements and bottlenecks, knowledge sharing, traceability, and process improvements were possible because the data for relevant metrics were available to support these objectives. At a higher abstraction level, these rationales evolve into the descriptive themes of Data availability, Plan-ning and bottleneck tracking, Information consistency, and Visibility. The last three rationales are the companies capitalizing on the data that is now available and accessible in the desired form due to the metrics. Therefore, further analysis of the four descriptive themes lead to the higher-order analytical theme of *Data Capitalization,* presented in Appendix C (https://goo.gl/Dtr5nx). Specifically, companies want to use the metrics data to create *awareness* about project activities by keeping relevant stakeholders in the loop, and improving transparency across the organization. Secondly, the metrics data could help companies identify bottlenecks in resource planning, testing, and review phase, enabling them to exercise *control* over these processes as a measure of improvement.

### 4.2 RQ2: What are the challenges faced by software-intensive companies in operationalizing the choice of metrics to assess their ASD processes?

From the GQM workshops, we identified some common and some unique challenges concerning operationalization of the elicited metrics. We categorized these challenges in three groups, as presented in the following table:

**Table 4.** Challenges in Operationalizing Metrics

| Challenge | Description |
| --- | --- |
| Lack of data or appropriate tools to produce that data | Development practices and processes at a company does not produce the data needed to measure a metric, or the company is not aware what data they can retrieve, or they are not using the tool(s) needed to measure a metric. |
| Existing process inhibiting change | Existing development process does not result in the data needed to measure the metric a company is interested in. |
| Difficulty in deriving actionable inputs | Data to measure a metric is available, but a company is uncertain about that metric's potential in providing actionable inputs. |

**Lack of data or appropriate tools to produce that data.** Lack of data and lack of appropriate tools are closely interlinked. For example, despite interest in several metrics, Company A did not have suitable tools to produce the data needed for these metrics, as reflected in the following quote: "It's not available, because we are not using Gerrit [in the case concerned] just yet". Similarly, for Company C, availability of data depended on how easily it could be retrieved, as highlighted in the comment: "…if we could track this, we could explain a lot of other phenomenon in the development…and this is never done. It's difficult to collect, but useful to know". Data unavailability can also stem from unawareness of what data could be retrieved from a certain tool. The following comment supports this claim: "We don't know what information in this specific tool is available".

**Existing process inhibiting change.** Specific development practices and processes can pose a problem in operationalizing some metrics. For instance, identification of bugs is not made explicit during sprints for Company A. To use metrics that require such information, company will need to change its development process, which is a challenge as indicated in the quote: "That's a process change. We should then change our process. Everybody write everything to JIRA, and I know that nobody will do this…This will waste developer's time hugely". Similarly, another case company believed that it is theoretically possible to measure the metric 'Ticket Size', but in practice, it would require an additional task on their part, which they do not encourage. The same challenge was identified even in case of assessing *Issues' Estimation Accuracy* process factor. This particular challenge could manifest in many ways like case companies do not have a formal practice in place to collect the data for the metrics, or the management is not interested in assessing this process factor, or the metric is not compatible with the development practice followed. Overall, in order to measure some of these metrics, a change in existing software development process or practice is required. However, such a change may run the risk of compromising the agile aspect of their software development process.

Unavailability of data can be seen as a consequence of lack of relevant tools that produce the needed data, or lack of supporting development process, or both. Analysis of the above-discussed two specific challenges produced the higher-order theme of *Data Unavailability*, as the common underlying challenge in metrics operationalizing. The analysis is available in Appendix D (https://goo.gl/KRUZBX). One of the common manifestations identified for this challenge is *process inertia*, which can be viewed as a condition where development process related aspects obstruct a case company from operationalizing their choice of metrics.

**Difficulty in deriving actionable inputs.** A company may have the right tool and the supporting development process to collect the necessary metric data. However, application of that data within a larger strategic context poses a challenge. Company B illustrated this challenge as follows: "…we have plenty of data and tools to collect metrics, but we have shortcomings for efficiently and smartly utilizing the collected data". A similar supporting inference was drawn from the following comment made by a stakeholder of Company C: "It has to add value to the measure…what I'd like to have is 'green' or 'red' light about my project. That's enough. I don't want hundreds of measures, curves, and pie-chart, and so on".

Analysis of the above challenge produced a higher-order theme of *Lack of Actionable Input* (in Appendix D - https://goo.gl/KRUZBX) representing the second challenge in metrics operationalization in our study. Essentially, case companies expect their chosen metrics to facilitate decision-making, or at least add enough value to stimulate actions toward improving their development process. Extraction of such actionable inputs from metrics is difficult, as expressed by one of the case companies.

## 5    Discussion

In this section, we discuss the multiple case study's results and address the two RQs. In comparison to existing literature, we find that the study's results reinforce and even extend existing knowledge, particularly in case of RQ2.

### 5.1    RQ1: What are the metrics software-intensive companies interested in to assess their ASD processes, and the rationale behind them?

We found that the case companies are interested in assessing processes related to implementation (*Issues' Velocity,* and *Development Speed)*, testing (*Testing Performance* and *Testing Status),* and planning (*Issues' Estimation Accuracy* and some metrics under *Issues' Velocity*). Research suggests that interest in such process factors indicates the need to plan and track sprints and projects [5]. *Issues' Velocity* metrics like 'No. of issues/ticket at the start of the sprint' and 'No. of issues/ticket at the end of the sprint' point to the need to assess sprint-planning capability. Similarly, by identifying bottlenecks by assessing *Delivery Performance* process factor, a company could learn if there is a need to plan their releases better. Such metrics suggest that agility imparted by ASD does not make a company immune to concerns associated with traditional software development [5]. Moreover, metrics assessing sprint or project velocity align with the third Agile principle of delivering working software in shorter cycles [21]. Similarly, factors like *Blocking Code* and *Quality Issues' Specification* reflect the Agile principle of continuous attention to technical excellence and good design [5, 21, 22]. Compliance to Agile principles indicates that the case companies selected metrics that stayed true to the tenets of ASD.

Variations in development context may not fully affect what process factors a company is interested in (4/10 factors are common), but it certainly influences how they are used. This is apparent in assessment of factors like *Testing Performance, Issues' Velocity, Code Quality, Testing Status,* and even *Blocking Code* to some extent. In addition to common metrics, case companies defined several custom metrics that aligned with their development context. Current literature [5] identifies company size and project characteristics as the development context aspects that can influence a company's choice of metrics, and our findings further reinforces this claim. For instance, Company D (small company) mostly preferred low-level metrics (specific measurement like 'No. of issues that are delayed') to high-level metrics (complex measurement like 'Average time to fix an error'). In addition, a typical project at Company D lasts for around four months. This further reinforces their interest in low-level metrics, as such metrics can help them gather insights at a lower granularity. The opposite was true for Company B (large company), where low-level process metrics proved to be inadequate at measuring process aspects for the large-size features the company implements. Furthermore, a typical project at Company B can take years, involving development of large features by a team of thousands. Hence, process insights at the lowest granular level may be insufficient at assessing development processes supporting development of large-scale product. Instead, higher-level metrics can provide the relevant insights to Company B in a condensed format. Existing research argues that company size and project characteristics determine adoption of ASD. Based on our findings, we argue that these determinants can also influence adoption of certain process metrics in ASD.

Rationales like identifying bottlenecks, tracking improvements, knowledge sharing find support in [5]. However, *Data Capitalization* is the overarching rationale that ties these disparate rationales together. Case companies want to use metrics to derive insights from the large amount of data produced by their systems and processes, and communicate this knowledge across the organization to enhance visibility (create *awareness*). Subsequently, the companies want to use this knowledge to exercise control in an effort to improve planning, track progress, and manage information integrity. The need to create awareness among stakeholders and use controlling measures to induce improvements is also reflected in current literature [23, 24]. Exercising control is in contrast to the tenets of Agile, but research argues that successful software companies tend to plan and estimate projects accurately [25], even in case of ASD [5].

### 5.2 RQ2: What are the challenges faced by the software-intensive companies in operationalizing the choice of metrics to assess their ASD processes?

In contrast to the fundamental rationale of data availability, data unavailability is one of the two fundamental challenges to obstruct metrics operationalization. *Process Inertia* was identified as a manifestation of this challenge. Ideally, to overcome such a challenge, an existing development process needs to be changed, but this is not always considered a feasible alternative. Such challenges are in line with the requirement that metrics in ASD should adhere to a company's development context [26, 27], be lightweight, and not hinder normal development activities [28]. The incompatibility between some of the process metrics chosen by the case companies and their development context appears to be the cause of process inertia, further translating into one of the two main challenges of *Data Unavailability*.

The second challenge of *Lack of Actionable Input* is supported by the concerns raised by some case companies that metrics should ideally reflect or stimulate actionable inputs, geared towards decision-making. The case companies expected their chosen metrics to help them in producing actionable inputs to support their decision-making. This philosophy resembles that of *actionable analytics*, where practitioners derive actionable inputs from the predictive capacity of the collected data [29]. In the absence of such information, a metric may not benefit a company in ways it expects it to. It is also required that a company be able to extract actionable input from a metric, and utilize it in an efficient and even strategic ways. However, the case companies were skeptical if use of metrics' would indeed lead to actionable inputs. Furthermore, they indicated that extracting actionable inputs from a metric can be a struggle.

## 6 Threats to Validity

Potential misinterpretation of GQM goal and the questions can be a threat to our study's construct validity. By explicating the GQM goal, questions, and having the findings and analyses validated by the case companies helped mitigate the threat to construct validity.

Internal validity concerns examination of causal relations free of influences unknown to the researcher [19]. The primary source of threat to our study's internal validity could be the selection of participants for the GQM workshops, as these could lead to elicitation of metrics that are irrelevant from ASD process standpoint. A company champion helped us identify stakeholders that are responsible for taking process related decisions in each case company, thereby mitigating the said threat. Next, two researchers executed and analyzed the data collected from the GQM workshops, which were shared with the case companies for feedback. This also helped mitigate threat to the study's internal validity.

Generalizability (external validity) of our multiple case study findings is limited only to the contexts of the four case companies that participated in the study. However, analysis and integration of other similar cases could extend our results, and companies with context similar to any of the case companies may find our findings applicable.

Reliability is related to the extent to which the data and the analysis are dependent on a researcher [19]. Multiple researchers participated in both data collection and data analysis, as a measure to enhance the study's reliability.

## 7 Conclusion and Future Work

Existing scientific literature focuses on the impact of using metrics in industrial ASD, but the associated rationales and challenges remain underexplored. Using the GQM approach, we conducted a multiple case study, involving four case companies, to address this research gap.

The case companies were interested in similar process factors like *Issues' Velocity, Testing Performance, Issues' Estimation Accuracy,* and *Code Quality.* Depending on individual requirements, case companies also wanted to measure exclusive process factors like *Development Speed, Delivery Performance,* and *External Quality.* Rationales such as data availability, tracking planning and bottlenecks, traceability, and knowledge sharing support the selection of metrics. However, being able to capitalize on the data to create awareness and exercise control over development processes appear to be the fundamental rationales. Data unavailability, a consequence of prevailing development context like limiting technical infrastructure or inhibiting development process, underlie several individual challenges that can obstruct metrics operationalization. For a company to extract actionable input from a metric to seek value addition or facilitate decision-making is another challenge that deserves further attention.

Our paper is part of the larger research project to help companies make data-driven (informed) decisions in Agile and rapid software development. The case companies are in the process of operationalizing the metrics reported in this study. Our future scientific studies will be about observing the influences these metrics have on ASD process at the case companies, and how they translate these metrics into actionable inputs.

**Acknowledgment.** This work is a result of the Q-Rapids Project, which has received funding from the European Union's Horizon 2020 research and innovation programme, under grant agreement No. 732253.